# WeChat Uptake of Chinese Scholarly Journals: An Analysis of CSSCI-indexed Journals

Ting Cong[1], Zhichao Fang[2], and Rodrigo Costas[3]

[1] *congting13@163.com*
University of Shanghai for Science & Technology, Dept of Publishing, Shanghai City (China)

[2] *z.fang@cwts.leidenuniv.nl*
Centre for Science and Technology Studies (CWTS), Leiden University, Leiden (the Netherlands)

[3] *rcostas@cwts.leidenuniv.nl*
Centre for Science and Technology Studies (CWTS), Leiden University, Leiden (the Netherlands)
DST-NRF Centre of Excellence in Scientometrics and Science, Technology and Innovation Policy, Stellenbosch University, Stellenbosch (South Africa)

**Abstract**

The study of how science is discussed and how scholarly actors interact on social media has increasingly become popular in the field of scientometrics in recent years. While most prior studies focused on research outputs discussed on global platforms, such as Twitter or Facebook, the presence of scholarly journals on local platforms was seldom studied, especially in the Chinese social media context. To fill this gap, this study investigates the uptake of WeChat (a Chinese social network app) by the Chinese scholarly journals indexed by the Chinese Social Sciences Citation Index (CSSCI). The results show that 65.3% of CSSCI-indexed journals have created WeChat public accounts and posted over 193 thousand WeChat posts in total. At the journal level, bibliometric indicators (e.g., citations, downloads, and journal impact factors) and WeChat indicators (e.g., clicks, likes, replies, and recommendations) are weakly correlated with each other, reinforcing the idea of fundamentally differentiated dimensions of indicators between bibliometrics and social media metrics. Results also show that journals with WeChat public accounts slightly outperform those without WeChat public accounts in terms of citation impact, suggesting that the WeChat presence of scientific journals is mostly positively associated with their citation impact.

## 1. Introduction

*1.1. Social media uptake by scholarly journals*

The study of how science is discussed on social media has increasingly become a popular topic in scientometrics in recent years. Social media are used by various stakeholders in the scientific community, including not only individual researchers, but also scholarly journals and publishers. Some scholarly journals have started to use Twitter to increase their online visibility and promote their published contents. However, when compared with social media usage by the general public, the social media uptake of scholarly journals is still lagging behind. For example, according to previous studies, scholarly journals included in the Web of Science (WoS) indexes Arts & Humanities Citation Index (A&HCI), Science Citation Index Expanded (SCIE), and Social Sciences Citation Index (SSCI) having Facebook accounts accounted only for 14.2%, 7.7%, and 7.2%, respectively (Zheng et al., 2019). It is therefore clear that the presence of scholarly journals on social media is still a relatively ongoing and underdeveloped activity.

*1.2. WeChat as a relevant local altmetric data source*

To date, altmetric data providers mostly track events and metrics from global social media platforms such as Twitter, Facebook, or Mendeley. These global social media platforms are also the data sources that have been most widely studied in the altmetric literature (Sugimoto et al., 2017). However, local social media platforms like WeChat - one of the most used social media platforms in China, which provides users with a platform for not only entertainment but also professional usage – has seldom been studied in altmetrics, although arguably it serves as a valuable tool for academics and academic organizations to publicly communicate about their research developments. In a 2019 *Springer Nature* survey, 94% of the 528 respondents in China indicated that they had used WeChat in a professional context (Nature Methods, 2020). Therefore, it can be argued that WeChat has the potential to become a relevant local altmetric data source.

WeChat was initially launched as an instant messaging app by Tencent in January 2011 and later evolved into a multipurpose app through its integration with a social media platform, a mobile payment system, and an e-commerce shopping application. According to Statista, by 2021, WeChat has over 1.24 billion monthly active users and accounts for 34% of China's total data traffic.[1] Besides, there are 1.48 million monthly active users of WeChat in the US.[2] With the growing number and internationalization of users, WeChat has become an important platform for science communication, offering the potential of mapping and evaluating the communication of science in the Chinese context (Xu, 2019). To help gain an insight into WeChat, we selected the two predominant Chinese social media platforms - Tencent WeChat and Sina Weibo - to briefly introduce their main services and their western equivalents. According to Table 1, while Weibo is similar to Twitter, WeChat has similar features to Facebook, although its closest western equivalent is possibly WhatsApp.

Table 1. Comparison of different social media tools in China

| Social media tools | Western equivalent | Main Sevrices |
|---|---|---|
| Tencent WeChat | Instant messaging app. Its closest western equivalent is WhatsApp, but when used as a social network app, WeChat has been liken to mobile Facebook. | - Find your friends more easily<br>- Start a free chat anytime and anywhere<br>- Share on Moments<br>- Free video and voice calls<br>- Hilarious and cute stickers |
| Sina Weibo | Microblog platform, analogous to Twitter. | - Powerful platform for public self-expression<br>- Any user can post a feed and attach multi-media and long-form content<br>- Any user can follow any other user and add comments to a feed while reposting<br>- A wide range of advertising and marketing solutions to organizations of all sizes |

WeChat Public Account (WPA) is one of the most important services embedded in WeChat, which was launched in August 2012. WPAs are often registered by organizations and companies, and are used to deliver posts to subscribers. The basic features of WPAs are broadcast messaging, auto-reply, content management and direct messaging. According to a list of academic WPAs,[3] which was launched by the Chinese research communication platform

---

[1] WeChat: active users worldwide. https://www.statista.com/statistics/255778/number-of-active-wechat-messenger-accounts/ (Accessed August 01, 2021).
[2] WeChat Statistics: https://99firms.com/blog/wechat-statistics/ (Accessed August 01, 2021).
[3] The ranking list of WeChat Public Accounts among 500 global scholarly journals (in Chinese):
https://www.linkresearcher.com/information/fa60d09a-0853-4796-9084-e4ec0b42b1d1 (Accessed August 01, 2021).

Linkresearcher (www.linkresearcher.com) and Impact Science from Cactus Communications, there are hundreds of Chinese and international scholarly journals that have established their own WPAs (including *Nature*, *Science*, and *Cell*).[4] For example, the *Nature Portfolio* WPA (see Figure 1A) published a total of 1,247 WeChat posts in the first quarter of 2021. Figure 1B shows an example of one of the Nature Portfolio's WeChat posts, while Figure 1C captures some user engagement metrics related to this WeChat post example, including the numbers of "readers", "likes", and "recommendations".

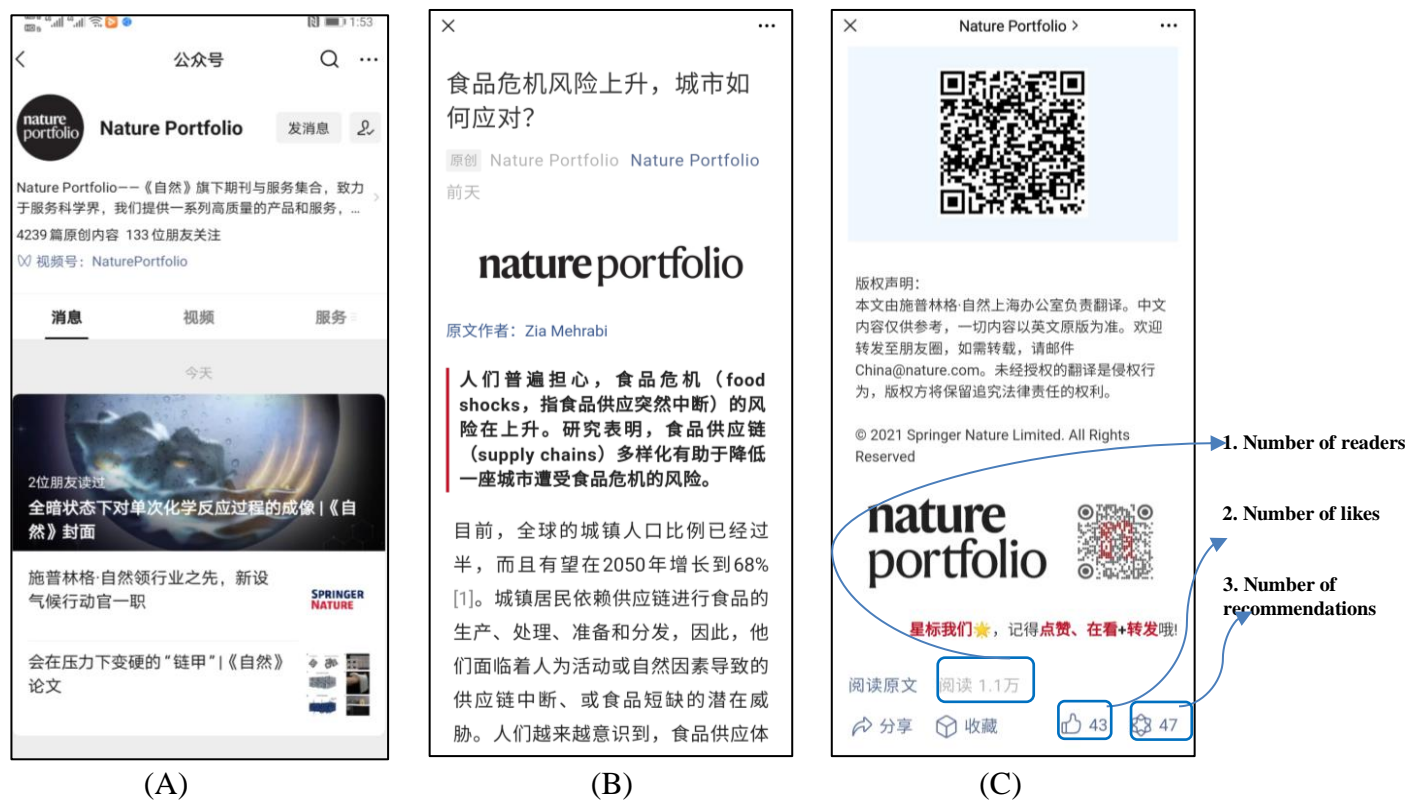

Figure 1. (A) The homepage of *Nature Portfolio*'s WPA, (B) a WeChat post example of the WPA, and (C) the user engagement metrics of the WeChat post example.

### 1.3. WeChat uptake by scholarly journals

Previous research about the social media uptake by scholarly journals had a focal point on global social media platforms, such as Twitter and Facebook (Kamel Boulos & Anderson, 2012; Nason et al., 2015). However, local social media platforms have seldom been investigated despite their potential value in understanding the online diffusion of scientific information by scholarly journals in local contexts. Given the popularity of WeChat in China, and the fact that it provides relevant analytical social media features (e.g., posts, readers, likes, recommendations), very similar to those of other global platforms like Twitter or Facebook, it can be fairly argued that WeChat is a strongly relevant platform for altmetric research in the Chinese context, albeit not yet explored in the existing literature on social media metrics.

Moreover, the vast majority of social media metrics research has focused on studying the presence of scientific papers on social media, whereas only a few studies analyzed the social media presence for scholarly journals. This study sets out to investigate the WeChat uptake by Chinese scholarly journals in the field of social sciences and humanities. Therefore, our object of study is scholarly journals instead of scientific articles. This is in line with the suggestion of Haustein et al (2016) that social media metrics do not need to be restricted to the analysis of the social media uptake of articles but can also consider the study of other academic agents active on social media, like scholarly journals.

---

[4] The global ranking of academic journals' impact on WeChat announced: https://cactusglobal.com/press/the-global-ranking-of-academic-journals-impact-on-wechat-announced/ (Accessed August 01, 2021).

*1.4. Objectives of this study*

Up to now, due to a series of barriers (mostly linguistic, infrastructural, and cultural), little is known about the WeChat uptake by Chinese scholarly journals and their use of WPAs. This study aims at filling this gap by exploring the extent to which Chinese scholarly journals are present on the WeChat platform. As such, the specific objectives of this research are:
- To investigate the overall WeChat uptake by Chinese scholarly journals from the field of social sciences and humanities;
- To investigate the relationship between traditional bibliometric indicators and WeChat-derived indicators at the journal level.

## 2. Literature Review

*2.1. Social media usage of scholarly journals*

Social media has been integrated into the system of scholarly communication in recent years (Sugimoto et al., 2017). Previous research on the uptake of social media in academia mainly focused on two main directions. One is the social media uptake by individual scholars (Rowlands et al., 2011; Tenopir et al., 2013), and the other is the social media uptake by scholarly organizations like scholarly journals and universities (Kortelainen & Katvala, 2012; Forkosh-Baruch & Hershkovitz, 2012).

Social media are mainly used by scholarly journals to announce their published articles and disseminate knowledge, complementing to traditional communication channels (Zheng et al., 2019). Most previous research focused on the uptake of social media by scholarly journals in specific subject areas. For example, Kortelainen and Katvala (2012) investigated 100 scholarly journals' websites to identify the use of social media tools by these journals (as reported in their websites) and to examine the attention data revealed by these social media tools. From another perspective, Boulos and Anderson (2014) selected the top 25 general medicine journals from the *Journal Citation Reports* to study their use of Facebook and Twitter, finding that 80% of the journals were present on Facebook, while 44% on Twitter. Zheng and his colleagues (2019) searched directly for the names of 13,826 journals indexed by the Web of Science using the search engines on Facebook and Twitter. It was found that the proportion of journals with Facebook accounts varied between 7.1% and 14.2% across disciplines. Karmakar et al. (2020) studied a set of 99,749 articles from 100 different journals, showing that journals with social media plugins integrated in their webpages (e.g., with the possibility of tweeting about articles or sharing them on Facebook directly from the journal website) got significantly higher social media mentions to their articles as compared to journals that did not provide such plugins.

*2.2. WeChat-related research*

Prior research on WeChat mainly aimed at understanding the motivations of users to use WeChat (Mao, 2014; Gan & Wang, 2015; Wang et al., 2015). For instance, Gan and Wang (2015) compared the adoption of two popular social media platforms in China: microblog and WeChat, and they found that social gratification represented the most important motivation to use WeChat. This view can be supported by the findings of Lien and Cao (2014) which confirmed that motivational factors, such as entertainment and sociality, impacted the attitudes toward WeChat usage.

WeChat offers useful and unique features beyond the traditional social media tools and can provide a range of services for research and information needs. Some researchers have investigated the use of WeChat by academic libraries. For instance, Xu and his colleagues (2015) explored the use of WeChat by the top 39 academic libraries in China, finding that almost one third of the academic libraries used WeChat as a promotional tool of collections and services. Wei and Yang's research (2017) showed that 84.6% of the "985" universities[5] opened up WeChat Libraries, which signified that WeChat Library has become one of the most important mobile service models of top university libraries in China. As to scholarly journals, the usage of WeChat is to improve their academic impact at both home and abroad. With a case study of Chinese Laser Press (CLP), Wang and his colleagues (2020) pointed out that connecting with researchers via WeChat helped CLP maintain relations with researchers throughout their career and contribute to the internationalization of the CLP journals and conferences. Besides, there have been several research evaluation studies relying on WeChat data. Zhao and Wei (2017) took WeChat as a new source of data to measure the impact of researchers' outputs on social media platforms. They proposed several indicators related to the user engagement with researchers' posts, such as readers and clicks to the "thumbs up" sign.

## 3. Data and Methods

*3.1. Data collection*

In this study, we use the list of scholarly journals indexed by the Chinese Social Sciences Citation Index (CSSCI) as our main dataset. CSSCI (http://cssci.nju.edu.cn/) is a selective citation index which covers Chinese scholarly journals in the field of Social Sciences and Humanities. The journals listed in the index are often regarded as the sources of the most cutting-edge research in the field of Social Sciences and Humanities in China. In the 2019-2020 edition of CSSCI, there were a total of 782 Chinese scholarly journals indexed in its core and expanded collections (including 568 journals from the core collection and 214 from the expanded collection).

For the 782 CSSCI-indexed journals, we harvested their bibliometric data from the database of China National Knowledge Infrastructure (https://www.cnki.net/, CNKI) between September 28, 2020 and October 8, 2020. CNKI is the most comprehensive gateway of knowledge of China with a complete collection of both academic journals and non-academic journals published in mainland China since 1915. Just to provide an idea of the relevance of CNKI in the Chinese context, there were over 120 million end users and more than 2.3 billion full-text downloads to scientific publications recorded only in 2019.[6] For each of the 782 journals in our dataset, we collected their total number of published papers (P), total number of citations received (TC), total number of downloads (TD), and journal impact factors (JIF). Figure 2 shows the bibliometric data of a CSSCI-indexed journal example displayed on the CNKI website.

---

[5] Project 985 is a boosting project to promote the Chinese higher education system as called for by former President Jiang Zemin at the 100th anniversary of Peking University on May 4, 1998. The objective is to develop, in cooperation with local government, several top universities to be world-leading.
[6] CNKI. https://oversea.cnki.net/index/Support/en/Introduction.html

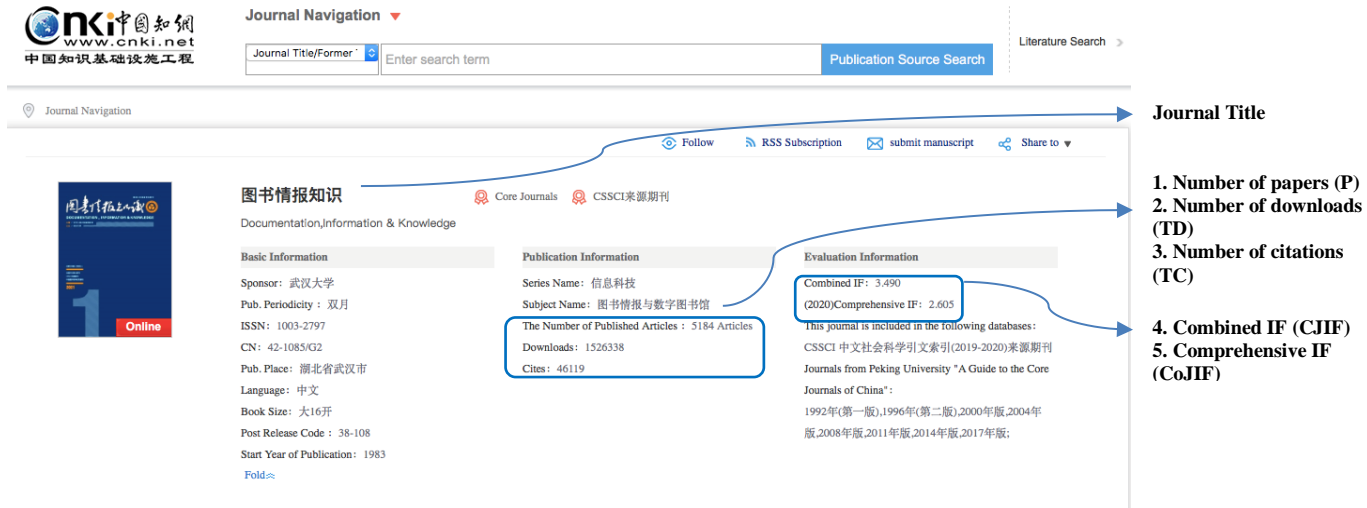

Figure 2. Bibliometric data of a journal example from the database of CNKI

To examine whether the CSSCI-indexed journals in our dataset have created WeChat public accounts, we manually searched the journals' titles by using the built-in search engine on WeChat. The profiles of the retrieved WeChat public accounts were scrutinized to ensure that they actually affiliate to the searched journals. Moreover, for those journals' WeChat public accounts with posts, we crawled the detailed information for each of their WeChat posts (e.g., post title, number of clicks, number of likes, and number of replies). The WeChat data collection was conducted during the period from October 10, 2020 to November 9, 2020.

### 3.2. Indicators and analytic approaches

Table 2 lists the bibliometric and WeChat indicators aggregated at the journal level and analyzed in this study. The bibliometric indicators of journals include number of papers (P), number of citations (TC), number of downloads (TD), combined IF[7] (CJIF), and comprehensive IF (CoJIF). The WeChat indicators of journals' WeChat public accounts include number of WeChat posts (TPosts), number of clicks (TClicks), number of likes (TLikes), number of recommendations (TRecomm), and number of replies (TReplies).

Table 2. Bibliometric and WeChat indicators aggregated at the journal level

| Category | Indicator | Abbr. | Concept |
|---|---|---|---|
| Bibliometric indicators | Number of papers | P | Total number of CNKI-indexed scholarly papers published by a journal until the year 2020. |
| | Number of citations | TC | Total number of citations (captured by CNKI) of all the papers published by a journal until the year 2020. |
| | Number of downloads | TD | Total number of downloads (captured by CNKI) of all the papers published by a journal until the year 2020. |
| | Combined IF | CJIF | Journal impact factor computed based on the citations contributed by **journal papers, conference papers and dissertations** indexed by CNKI from 2018-2019. |
| | Comprehensive IF | CoJIF | Journal impact factor computed based on the citations contributed by **only journal papers** indexed by CNKI from 2018-2019. |

---

[7] Both "combined IF" – combined impact factor – and "comprehensive IF" – comprehensive impact factor – are denominations of the indicators used by CNKI.
6

| Category | Indicator | Abbr. | Concept |
|---|---|---|---|
| WeChat indicators | Number of WeChat posts | TPosts | Total number of posts posted by a journal on its own WeChat public account until the data collection period. |
| | Number of clicks | TClicks | Total number of clicks received by all WeChat posts posted by the WeChat public account of a journal until the data collection period. |
| | Number of likes | TLikes | Total number of likes received by all WeChat posts posted by the WeChat public account of a journal until the data collection period. |
| | Number of recommendations | TRecomm | Total number of recommendations (by clicking "I'm reading it") received by all WeChat posts posted by the WeChat public account of a journal until the data collection period. |
| | Number of replies | TReplies | Total number of replies received by all WeChat posts posted by the WeChat public account of a journal until the data collection period. |

Figure 3 illustrates the research workflow of this study. This study starts with the collection of both WeChat and bibliometric data for the CSSCI-indexed journals, followed by aggregating WeChat data at the journal level. Finally, this study performs three types of analyses: general descriptive statistics, correlation and factor analysis performed on the IBM SPSS 25, and comparative analysis of scholarly impact between journals with and without WeChat public accounts created.

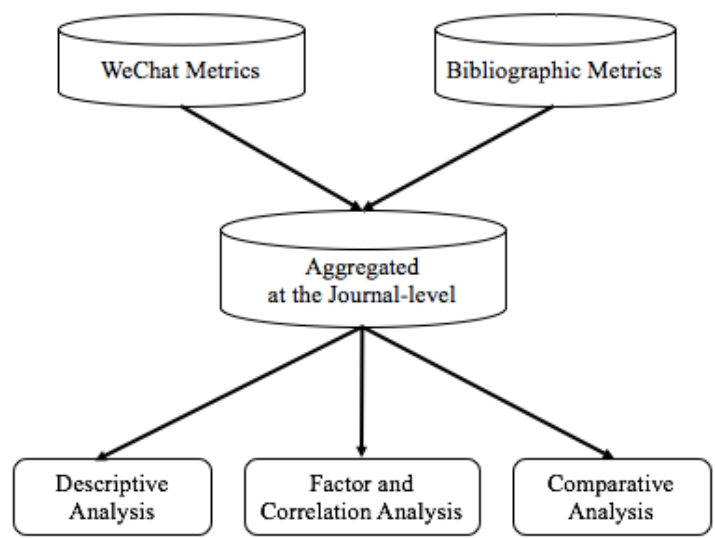

Figure 3. Research workflow

## 4. Results

### 4.1. WeChat uptake by CSSCI-indexed journals

Among the 782 CSSCI-indexed journals, 511 of them have WeChat public accounts (accounting for 65.3%). Table 3 shows the overall descriptive statistics of the five bibliometric indicators of the 782 CSSCI-indexed journals. There are 6 journals that have created WeChat public accounts but have not posted any WeChat post until the data collection. Therefore, we collected WeChat activity indicators for a total of 505 journals. Table 4 presents the descriptive statistics of the five WeChat indicators. The total number of WeChat posts produced by the 505 CSSCI journals is 193,367. These posts have received a total of 272.9 million clicks and 1.2 million likes, respectively. These numbers indicate the large amounts of interactions taking place around the content posted by scholarly journals on WeChat. Another important aspect that needs to be highlighted is the strong skewness of WeChat indicators. This can be demonstrated by the percentiles in Table 4, except total number of WeChat posts, with the mean of all indicators always higher than P75 (Quartile 3 (Q3)).

Table 3. Descriptive statistics of the bibliometric indicators (N=782 journals)

| *Indicator* | *Sum* | *Min* | *Max* | *Median* | *Mean* | *SD* |
|---|---|---|---|---|---|---|
| P | 4,436,708 | 131 | 39,541 | 4701 | 5673.54 | 4,298.627 |
| TC | 41,725,554 | 225 | 773,588 | 36839 | 53,357.49 | 58,398.515 |
| TD | 1,647,285,229 | 36,325 | 17,108,627 | 1,699,832 | 2,106,502.85 | 1,728,224.96 |
| CJIF | | 0.033 | 13.061 | 1.472 | 1.901 | 1.648 |
| CoJIF | | 0.033 | 8.436 | 0.838 | 1.118 | 1.016 |

Table 4. Descriptive statistics of the WeChat indicators (N=505 journals)

| *Indicator* | *Sum* | *Min* | *Max* | *Q1* | *Median* | *Q3* | *Mean* | *SD* |
|---|---|---|---|---|---|---|---|---|
| TPosts | 193,367 | 1 | 7,133 | 69 | 183 | 420.5 | 382.9 | 721.9 |
| TClicks | 272,923,439 | 78 | 42,621,361 | 33,904 | 107,585 | 301,034.5 | 540,442.5 | 2,363,561.6 |
| TLikes | 1,234,295 | 0 | 410,368 | 108.5 | 447 | 1,185.5 | 2,444.1 | 19,005.1 |
| TRecomm | 1,080,552 | 0 | 92,458 | 105 | 499 | 1,433 | 2,139.7 | 7,243.9 |
| TReplies | 329,240 | 0 | 78,595 | 1 | 42 | 223.5 | 652.0 | 4,063.4 |

*4.2. Factor analysis and correlation analysis of bibliometric and WeChat indicators*

In this section we focus on the factor and correlation analysis of bibliometric and WeChat indicators. Table 5 shows the results of factor analysis of the analyzed indicators aggregated at the journal level. Using principle component analysis (PCA), three main components were found. The first component is related to WeChat indicators, containing all WeChat indicators extracted from the WeChat platform. The second component is related to scholarly impact metrics, such as journal impact factors and number of citations (downloads also have a substantial relationship with this component). The third component is related to overall academic activity metrics, including the number of published papers and number of downloads recorded by CNKI database (in this component citations also play an important role).

Table 5. Factor analysis of the variables (N=505)

| Variable | Component | | |
|---|---|---|---|
| | 1 | 2 | 3 |
| TClicks | **0.980** | 0.030 | -0.015 |
| TReplies | **0.969** | -0.014 | -0.040 |
| TRecomm | **0.928** | -0.007 | 0.022 |
| TLikes | **0.910** | 0.009 | -0.055 |
| TPosts | **0.666** | -0.001 | 0.077 |
| CJIF | 0.018 | **0.987** | -0.148 |
| CoJIF | 0.004 | **0.979** | -0.154 |
| TC | -0.014 | **0.752** | *0.449* |
| P | 0.049 | -0.178 | **0.935** |
| TD | -0.019 | *0.592* | **0.652** |

Note: Principle Component Analysis (PCA) used as extraction method. Oblimin oblique rotation (pattern matrix, rotation converged in 9 iterations). Factor loadings higher than 0.400 are highlighted in italic, and those higher than 0.600 are highlighted in bold.

To further examine the relationships between WeChat indicators and bibliometric indicators, Spearman correlation analyses were conducted as shown in Figure 5 (N=505). Within both bibliometric indicators and WeChat indicators, indicators are generally moderately to strongly correlated. However, the correlations between bibliometric and WeChat indicators are generally weak or negligible. The correlation results stay consistent even if taking into account the journals without WeChat public accounts and assigning their WeChat indicators with zero values for analysis (N=782, see Figure 9 in Appendix). Both factor analysis and Spearman correlation analysis suggest that bibliometric and WeChat indicators capture substantially different types of interactions and impact.

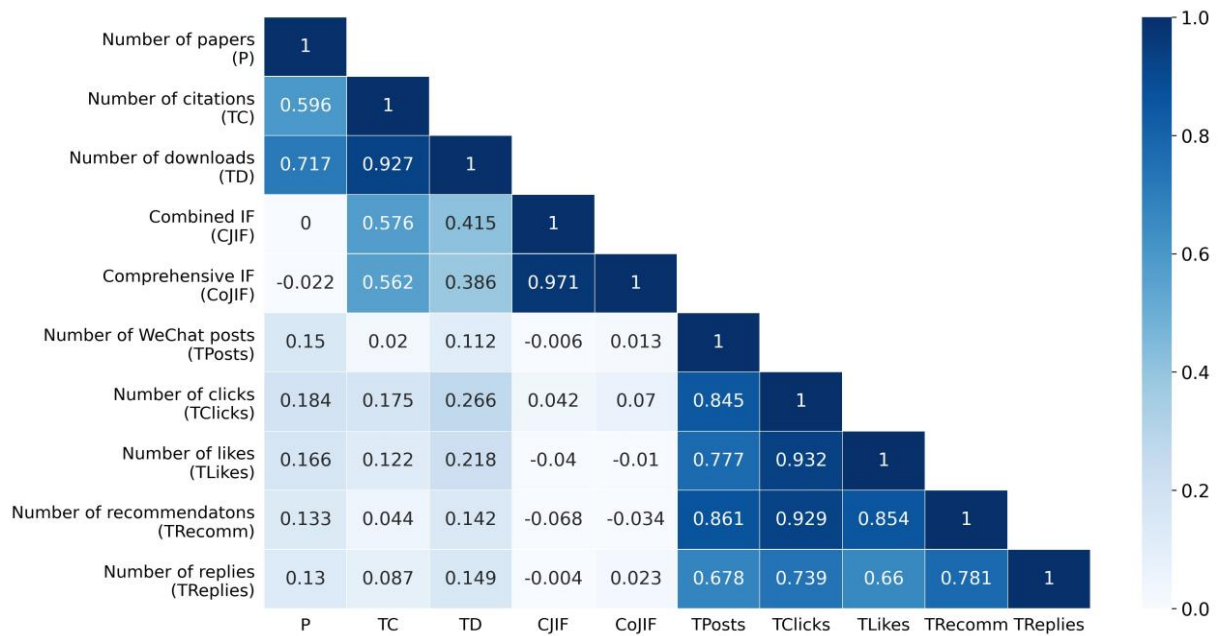

Figure 5. Spearman correlation analyses among bibliometric and WeChat indicators (N=505)

### 4.3. Comparison of bibliometric indicators between journals with and without WeChat public accounts

As there are only weak correlations between bibliometric indicators and WeChat indicators, in this section we explore whether those journals with WeChat public accounts are associated with somehow higher values of bibliometric indicators. Figure 6 plots the bibliometric indicators of journals with and without WeChat public accounts from the perspectives of (A) number of published papers, citations, downloads, and (B) the two journal impact factors. In general, journals with WeChat public accounts are associated with slightly higher bibliometric scores than those without WeChat public accounts.

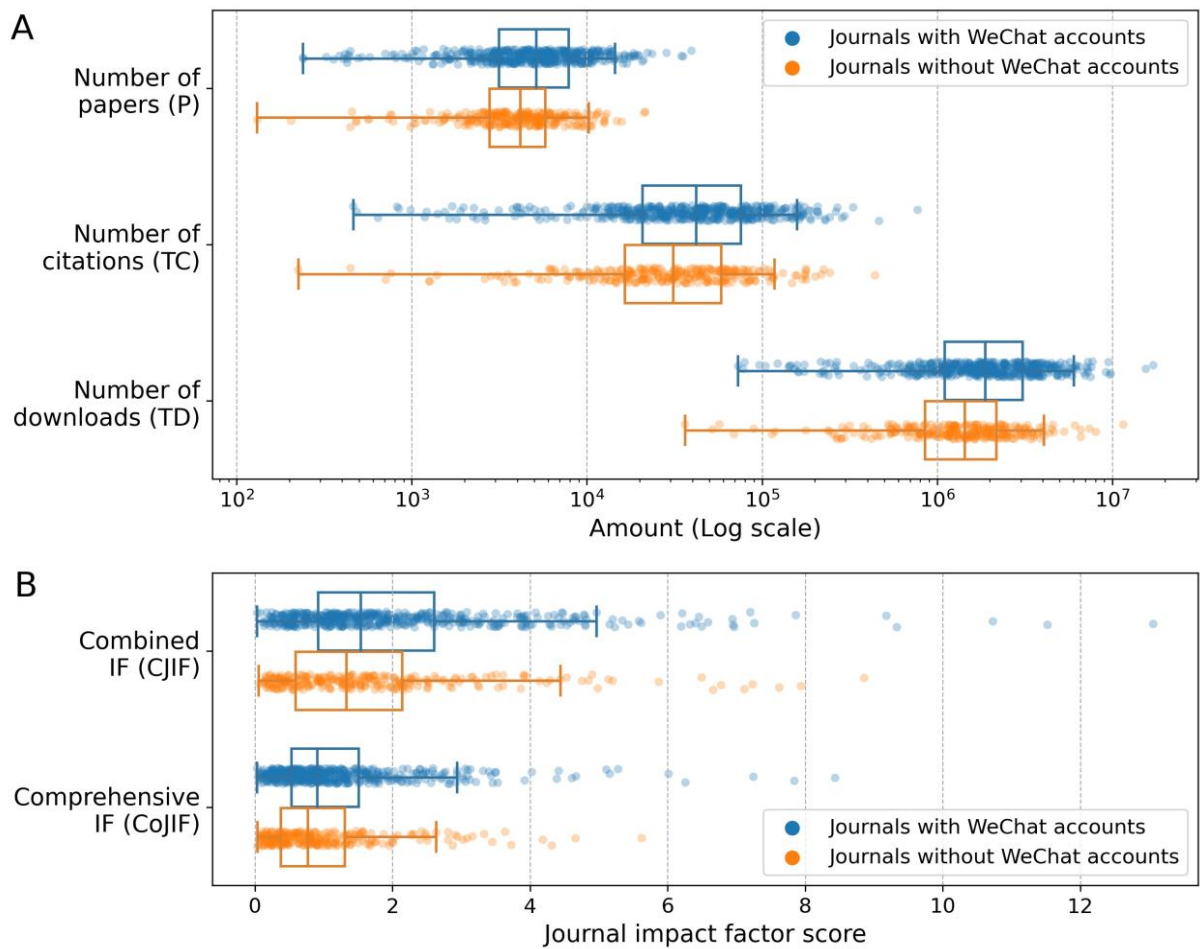

Figure 6. Bibliometric indicators of journals with (N=511) and without (N=271) WeChat public accounts

From the perspective of the user engagement metrics (i.e., clicks, replies, recommendations, likes) with the WeChat posts of scholarly journals, we focus on the total number of WeChat clicks (TClicks) as a representative of WeChat activity to classify scholarly journals with WeChat posts available (N=505) into two groups:
1) Journals with TClicks equal to or higher than the overall median value of TClicks for all of the journals analyzed (as per Table 4, TClicks ≥107,585).
2) Journals with TClicks below the overall median (as per Table 4, TClicks < 107,585).

Figure 7 compares the distribution of the five bibliometric indicators between these two groups. Overall, journals with their WeChat posts receiving larger amounts of clicks tend to also show relatively higher bibliometric scores, with the only exception of CJIF, with a lower median of those highly clicked journals. These results overall suggest a sort of positive but weak association between WeChat clicks and bibliometric impact indicators as already hinted in Figure 5, although no causal relationship can be suggested.

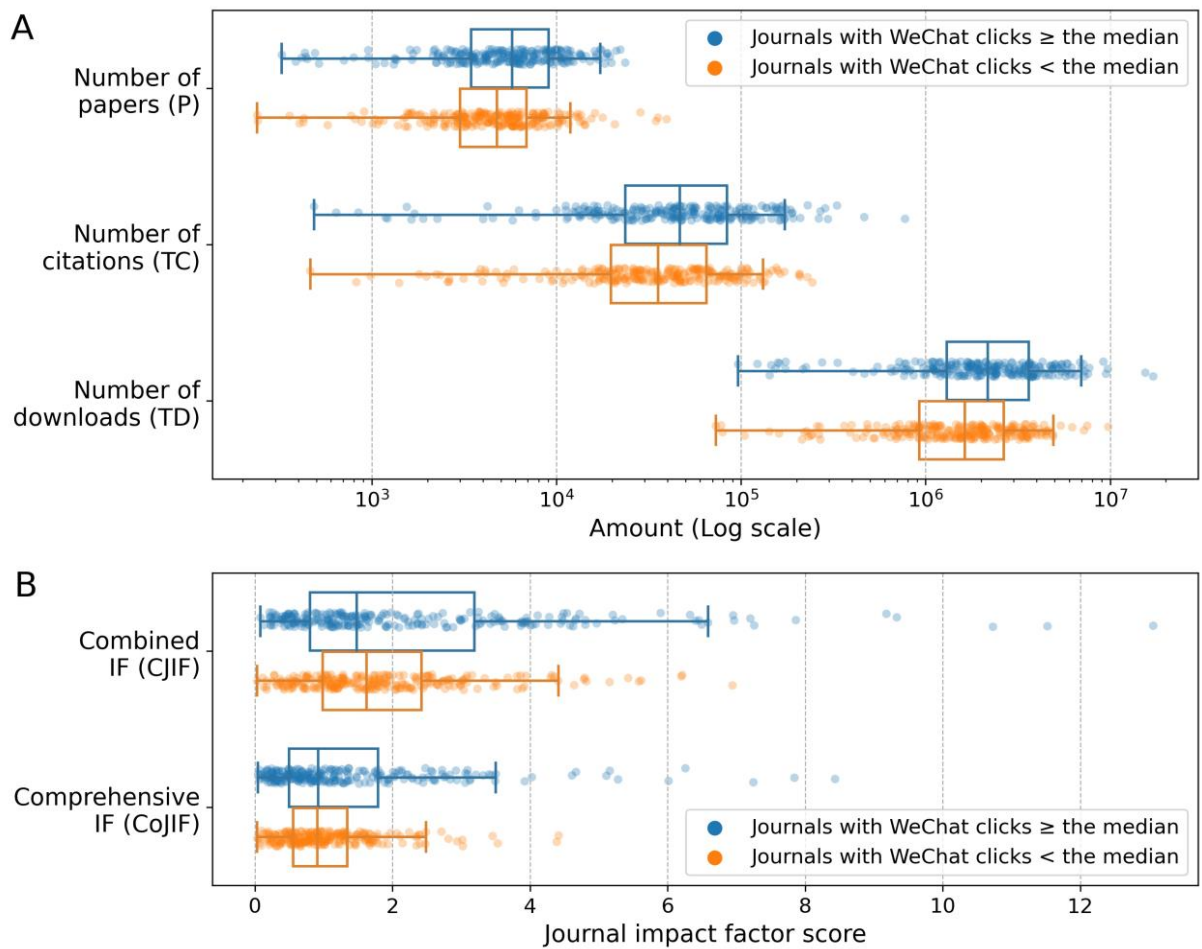

Figure 7. Bibliometric indicators of journals with number of WeChat click numbers not below the median and below the median (N=505)

*4.4. Comparison of WeChat indicators of journals across disciplines*

In order to compare the WeChat activities of journals from different disciplines, we classify the set of 505 CSSCI journals with WeChat posts available into three main subject categories with reference to the discipline list of Doctor and Master Degree Program[8] in China: *Art and Humanities category*, *Applied Social Science category*, and *Comprehensive category* (Table 6).

Table 6. Three types of subject categories among CSSCI-indexed journals (N = 505)

| Main subject category | Secondary subject category | Number of journals | Percentage |
|---|---|---|---|
| Art and Humanities category | Archelogy, Art, Chinese literature, Foreign Literature, Ethnology and Culturology, History, Linguistics, Philosophy, Religious Studies | 105 | 20.8% |
| Applied Social Science category | Economics, Education, Human Economic Geography, Journalism and Communication, Law, Library and Information Science, Management Science, Marxism, Politics, Psychology, Sociology, Sports Science, Statistics | 282 | 55.8% |
| Comprehensive category | Comprehensive Social Science, University Journal | 118 | 23.4% |

---

[8] http://www.cdgdc.edu.cn/xwyyjsjyxx/sy/glmd/264462.shtml (In Chinese, accessed August 01, 2021).

Figure 8 depicts the distribution of total number of WeChat posts, clicks, likes, recommendations, and replies across the three different subject categories of journals. Overall, journals from the three subject categories exhibit a quite similar degree of activity in posting WeChat posts (TPosts). At most through the lens of the medians displayed in the boxplots, it can be highlighted that WeChat posts from the WPAs owned by journals from the Art and Humanities category and the Applied Social Science category achieved slightly higher levels of user engagement (i.e., TClicks, TLikes, TRecomm, and TReplies) in contrast to journals in the Comprehensive category. However, from the perspective of bibliometric indicators (see Figure 10 in Appendix), journals in the Applied Social Science category and the Comprehensive category exhibit higher values than the Art and Humanities category, implying the different disciplinary biases between WeChat activities and academic activities.

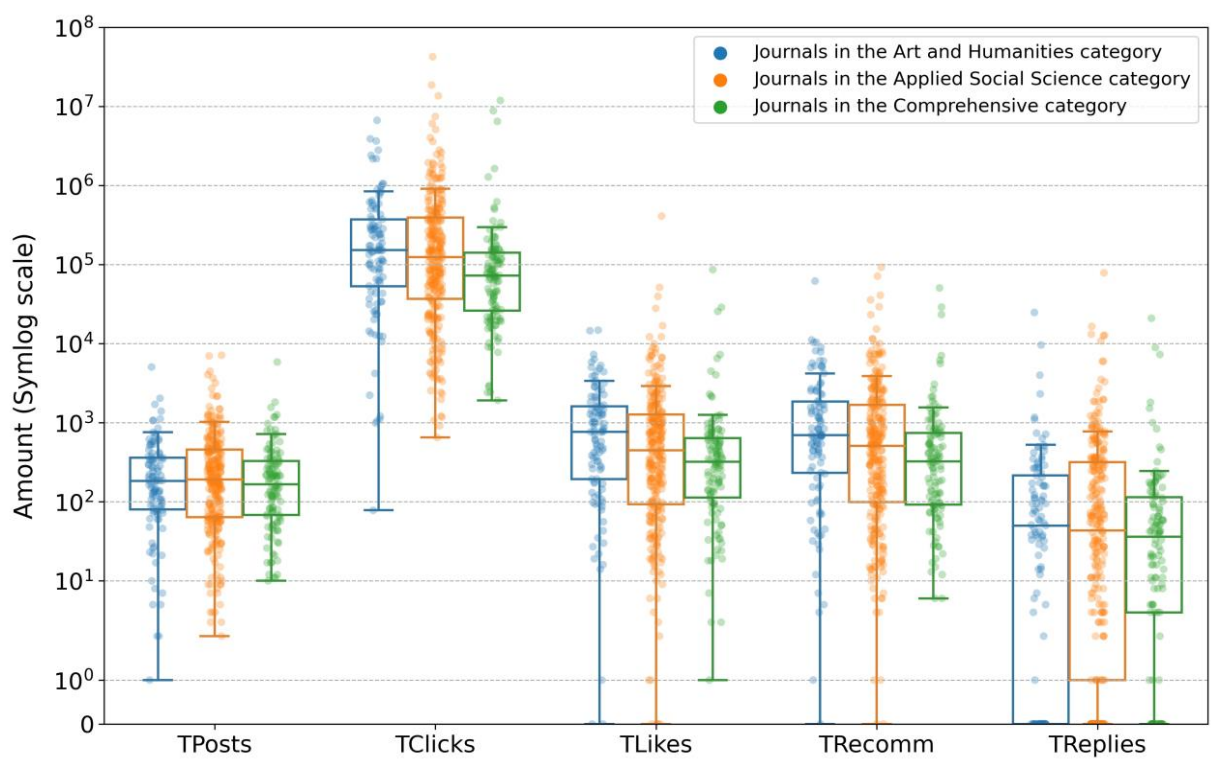

Figure 8. WeChat indicators of journals across the three subject categories (N=505)

## 5. Discussion and conclusion

This study empirically explored the presence of CSSCI-indexed journals on the WeChat platform. As such, it represents the first proof of concept of how also local social media sources can be incorporated in the altmetric study of also local scholarly journals. From a more conceptual point of view, following the classification by Yu et al. (2017), it can be argued that this study represents a type-C study, focusing on "local research discussed in local platforms". The findings of this research contribute to a better understanding of whether Chinese scholarly journals use local social media platforms to reach broader audiences. Given the broad uptake of WeChat by Chinese scholarly journals as well as the extensive interactions around WeChat posts by WeChat users, the findings of this study shed light on the importance of incorporating local sources in the study of social media metrics across countries.

*5.1. Overall uptake of WeChat by CSSCI-indexed journals*

The uptake of WeChat by CSSCI-indexed journals is rather high, with more than 65% of the journals analyzed having WeChat public accounts. In contrast, according to Zheng et al. (2019), only 14.2% of A&HCI-indexed journals and 7.2% of SSCI-indexed journals have Facebook[9] accounts. In the light of other previous studies, we can also confirm an increase in the number of CSSCI-indexed journals that have created WPAs, increasing from 53.1% reported by Zhang et al. (2018) to 65.3% in this study. This conclusion supports once again the relevance of WeChat as an important platform in the Chinese altmetric context, and more fundamentally supports previous calls for the consideration of local perspectives in altmetrics (Zahedi, 2017; Liu et al., 2020).

*5.2. Lack of correlation between WeChat indicators and bibliometrics*
The negligible correlations found between bibliometric and WeChat indicators confirmed once again the existence of two fundamentally differentiated dimensions of indicators. A dimension of indicators related to journal performance from a bibliometric point of view, and another dimension of indicators related to the WeChat uptake and WeChat engagement metrics. These results are in line with previous studies which reported the weak correlations between citations and most social media metrics (e.g., Twitter mentions, Facebook mentions, and blog citations) at the scientific paper level (Costas et al., 2015; Haustein et al., 2015). The results corroborate that WeChat indicators have a strong *social media focus*, rather than a *scholarly focus* (Wouters et al., 2019), supporting the idea that local social media metrics may share fundamental traits with global social media metrics (e.g., lack of correlation with bibliometrics, highly skewed distributions, strong correlations within social media metrics) (Díaz-Faes et al., 2019; Costas et al., 2016; Haunschild & Bornmann, 2017; Thelwall, 2021). This observation, made for the first time on a local social media platform, calls again for further research on the fundamental aspects of social media metrics that are independent of their global or local orientation. As such, the potential existence of such fundamental aspects could pave the way towards generalized social media metrics theories and frameworks (e.g., like the framework of "heterogeneous couplings" recently proposed by Costas et al., 2020), which could be relevant to explain and work both with global and local social media metrics.

*5.3. Association between WeChat uptake and bibliometric indicators*
According to previous studies, scholarly journals with a large number of followers on Facebook or Twitter tend to have a high JIF (Cosco, 2015; Raamkumar et al., 2018; Zheng et al., 2019). Even though there was no strong relation with the JIF for urological journals (Nason et al., 2015), it is found that the Twitter or Facebook presence of journals to be positively associated with the JIF (Kelly et al., 2016). From Cosco's research (2015) on Medical journals in the Twittersphere, the size of a general medical journals' Twitter following is strongly linked to its impact factor and citations, suggesting that high quality research tend to receive more social media attention. Although as discussed earlier there is only a weak relationship between bibliometric indicators and WeChat indicators, journals with WPAs tend to perform slightly better in terms of all of the analyzed bibliometric indicators. Since a causal relationship cannot be established (i.e., that being in WeChat does lead to higher bibliometric performance), we may, however, argue that being on the platform also does not seem to bring any bibliometric disadvantage. This means that, at a minimum, those journals that have opted for creating and maintaining a WeChat account do not exhibit a lower bibliometric performance than those who decided not to create one. It remains anyway an open question for future studies whether there are direct academic benefits (e.g., attraction of local authors, more diverse readers, citations) or societal benefits (e.g., larger societal awareness of the journals, more engagement with the journals, higher presence on social media debates) derived from the presence of scientific journals on WeChat or other social media platforms.

---

[9] We argue that Facebook can be seen as the western social media platform conceptually closer to WeChat. In any case, Zheng et al. (2019) also analyzed the Twitter presence of A&HCI and SSCI-indexed journals, finding only 9% and 7.7% of the journals with Twitter accounts, both numbers clearly below the values found in our study.

*5.4. Disciplinary differences*

To compare how journals from different disciplines vary in the attention towards their WPA posts, we divided the CSSCI-indexed journals into three main subject categories, namely, the Art and Humanities category, the Applied Social Science category, and the Comprehensive category. Our results show that journals from the three disciplines perform rather similar in their WeChat performance. Our results are somehow different from the observations by Costas et al. (2015) and Hammarfelt (2014) that scholarly outputs in Humanities exhibit a comparatively lower presence on global social media. According to our results, Chinese scholarly journals in the subject category of Art and Humanities are generally as active as the journals related to Applied Social Sciences in not only posting WeChat posts but also triggering user engagement on WeChat. Local social media and their users may exhibit a different disciplinary bias in the dissemination of local research outputs than global social media, which is also relevant to future research.

*5.5. Limitations and further studies*

Finally, as an exploratory research, this study also has several limitations. First, it only provides an overall descriptive picture of the WeChat uptake by CSSCI-indexed journals, thus being restricted to certain fields and the actual set of journals. In future research, we plan to explore in more detail the different activities on WeChat of other Chinese scholarly journals, as well as more advanced disciplinary and content studies of the WeChat posts. Second, the demographics of WPAs (e.g., the number of subscriptions) were not collected and analyzed in this study. Given the potential influence of the demographics of WPAs on the WeChat activities (e.g., on the number of clicks received by WeChat posts, or the ratio of likes or clicks per user subscribed[10]), it would be important to include such demographic variables in future research. Third, our study has shown how a non-negligible amount of online social media interactions (e.g., posts, clicks, likes, etc.) related to scholarly journals are recorded by a local social media platform like WeChat. This reinforces the importance of incorporating local sources in the study of social media interactions around science, since just focusing on western or global tools may miss out broader audiences like the Chinese social media community. These are all topics that deserve further attention in future research. In line with this, it is important to note that an important challenge for the reliable incorporation of WeChat as a platform for altmetric research is that WeChat does not have a public API that can be used to systematically collect data. Thus, our data collection for this study needed to be in a substantial part performed using ad hoc tools and manual processing of the data. It would therefore be desirable that WeChat would open an API that enables the type of research suggested in this study, enabling larger communities of researchers to engage in WeChat-based social media studies of science.

**Acknowledgements**

This paper is supported by National Key R&D Program of China (2021YFF0900400). Rodrigo Costas is partially funded by the South African DST-NRF Centre of Excellence in Scientometrics and Science, Technology and Innovation Policy (SciSTIP). The paper is a substantially extended version of a conference poster paper accepted at the 18[th] International

---

[10] There are several ways WeChat users can find and follow WeChat official accounts. The most common way of acquiring new followers on WeChat is through WeChat Moments (the WeChat equivalent of the Facebook timeline). Upon clicking on the article, users can access the account page by clicking the name of the account at the top of the article.

Conference on Scientometrics and Informetrics (ISSI 2021), celebrated the 12-15 July 2021 (https://issi2021.org/).

**Competing interest**
One of the authors (Rodrigo Costas) is a member of the Distinguished Reviewers Board of the journal Scientometrics.

# Appendix

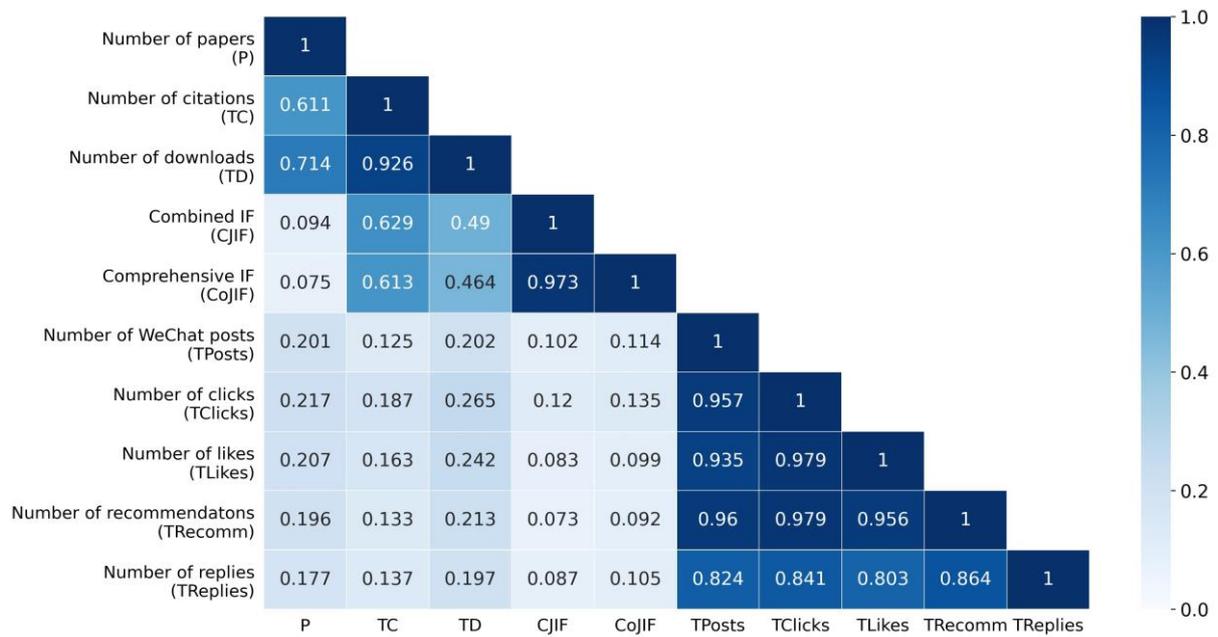

Figure 9. Spearman correlation analyses among bibliometric and WeChat indicators (N=782)

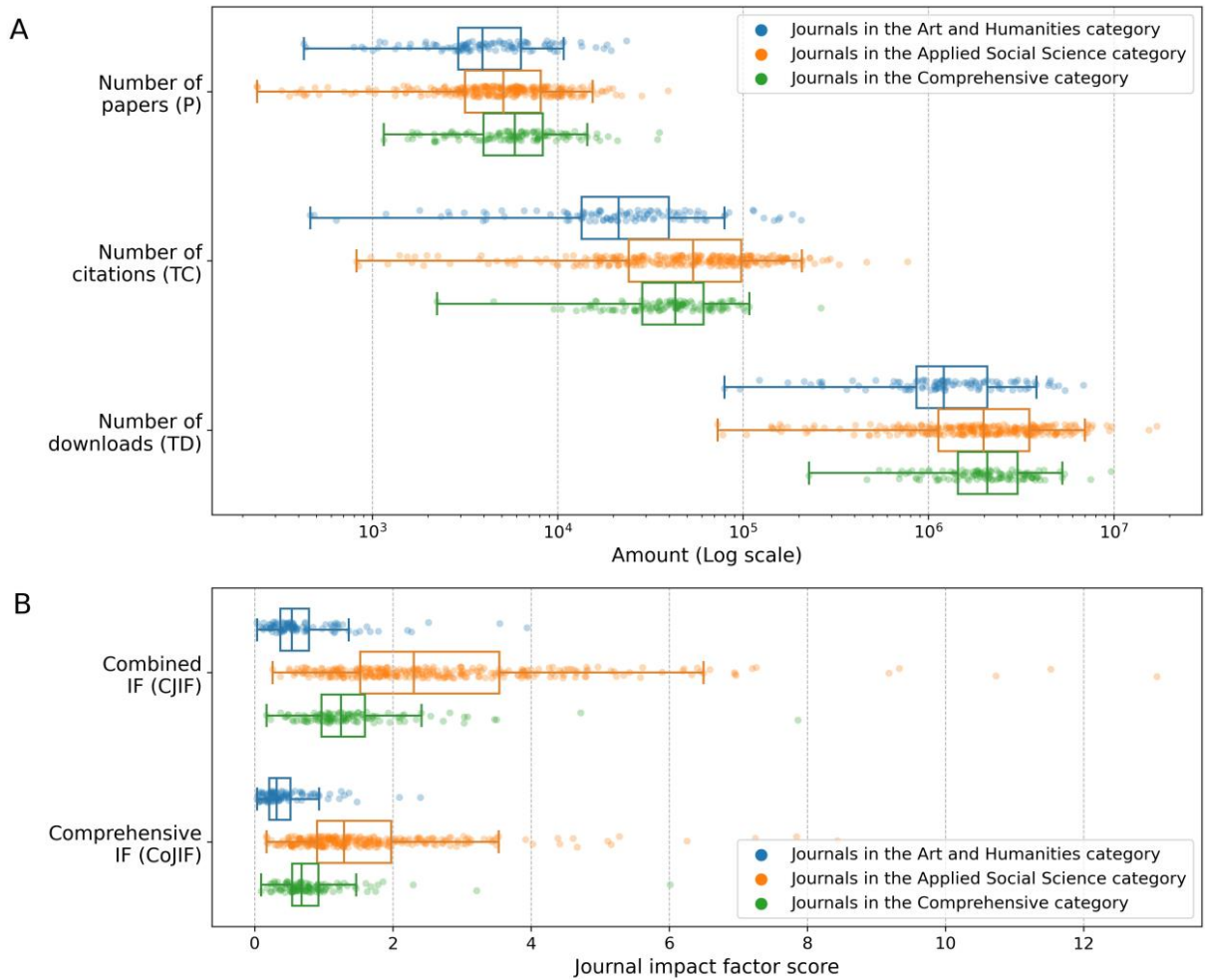

Figure 10. Bibliometric indicators of journals across the three subject categories (N=505)